\documentclass{aa}  

\usepackage[varg]{txfonts}
\usepackage{color}
\usepackage{longtable}
\usepackage{natbib}
\usepackage{graphics,graphicx,url}

\newcommand{\iras}{IRAS~17020+4544}

\newcommand{\xmm}{XMM-{\it Newton}}
\newcommand{\ergs}{$\mathrm{erg \ s}^{-1}$}
\newcommand{\whz}{$\mathrm{W \ Hz}^{-1}$}
\newcommand{\kms}{$\mathrm{km \ s}^{-1}$}

\bibpunct{(}{)}{;}{a}{}{,} 

\begin{document}

\title{Coexistence of a non thermal jet and a complex ultra fast  X-ray outflow in a moderately luminous AGN}

\author{M.~Giroletti\inst{1}\fnmsep\thanks{Email: giroletti@ira.inaf.it}, 
  F.~Panessa\inst{2}, 
  A.\,L.~Longinotti\inst{3,},
  Y.~Krongold\inst{4},
  M.~Guainazzi\inst{5,6},
  E.~Costantini\inst{7},
 M.~Santos-Lleo\inst{5} 
}

\institute{INAF Istituto di Radioastronomia, via Gobetti 101, 40129 Bologna, Italy \and
INAF Istituto di Astrofisica e Planetologia Spaziali di Roma, Via del Fosso del Cavaliere 100, 00133 Roma, Italy \and
CONACYT - Instituto Nacional de Astrof\'isica, \'Optica y Electr\'onica, Luis E. Erro 1, Tonantzintla, Puebla, M\'exico, C.P. 72840 \and
Instituto de Astronomia, Universidad Nacional Autonoma de Mexico, Apartado Postal 70264, 04510 Mexico D.F., Mexico \and
ESAC, P.O. Box, 78 E-28691 Villanueva de la Ca\~nada, Madrid, Spain \and
Institute of Space and Astronautical Science, 3-1-1 Yoshinodai, Chuo-ku, Sagamihara, Kanagawa, Japan \and
SRON Netherlands Institute for Space Research, Sorbonnelaan 2, 3584 CA Utrecht, The Netherlands
}

\date{Received ; accepted }

  \abstract
{Recent XMM-Newton observations have revealed that \iras\ is  a very unusual example  of  black hole wind-produced feedback by a moderately luminous AGN in a spiral galaxy.}
{Since the source is known for being a radio emitter, we investigated about the presence and the properties of a non-thermal component.}
{We observed \iras\ with the Very Long Baseline Array at 5, 8, 15, and 24 GHz within a month of the 2014 XMM-Newton observations. We further analysed archival data taken in 2000 and 2012.}
{We detect the source at 5 GHz and on short baselines at 8 GHz. At 15 and 24 GHz, the source is below our baseline sensitivity for fringe fitting, indicating the lack of prominent compact features. The morphology is that of an asymmetric double, with significant diffuse emission. The spectrum between 5 and 8 GHz is rather steep ($S(\nu)\sim\nu^{-(1.0\pm0.2)}$). Our re-analysis of the archival data at 5 and 8 GHz provides results consistent with the new observations, suggesting that flux density and structural variability are not important in this source. We put a limit on the separation speed between the main components of $<0.06c$.}
{\iras\ shows interesting features of several classes of objects: its properties are typical of compact steep spectrum sources, low power compact sources, radio-emitting narrow line Seyfert 1 galaxies. However, it can not be classified in any of these categories, remaining so far a one-of-a-kind object.}


  \authorrunning{M.\ Giroletti et al.}
  \titlerunning{Radio properties of \iras}

   \maketitle

\begin{figure*}
\includegraphics[width=0.96\columnwidth]{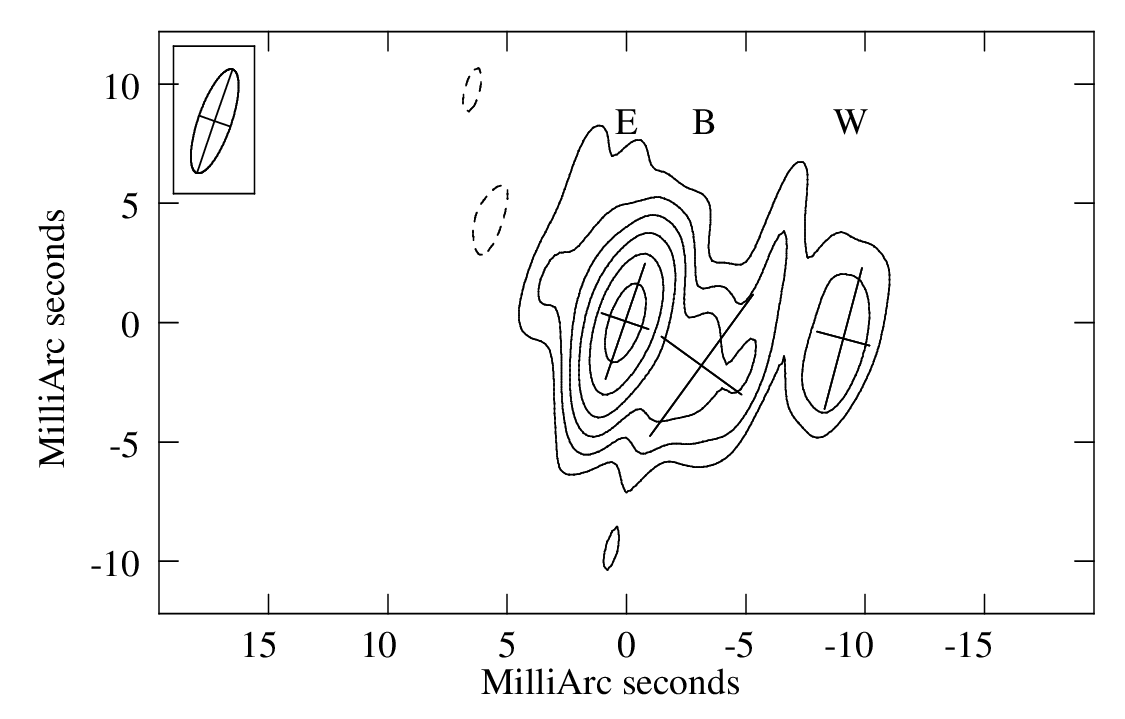}
\includegraphics[width=0.96\columnwidth]{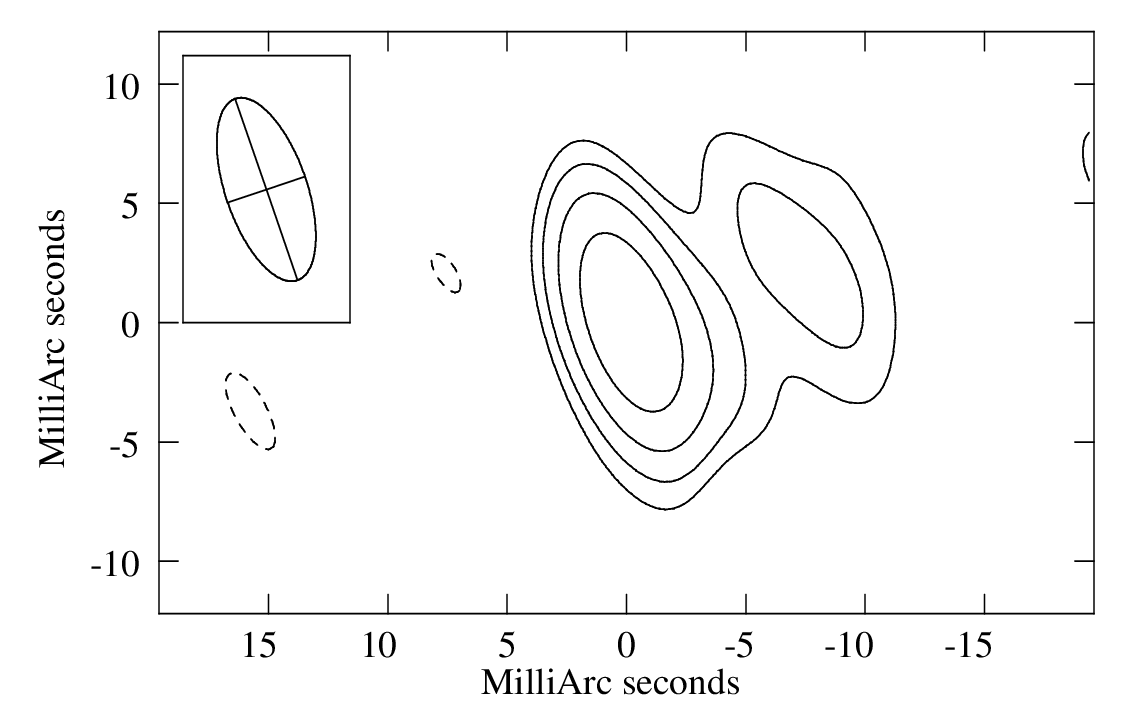}
\caption{VLBA images of \iras\ on 2014 February 12 at 5 GHz (left hand side panel) and 8.4 GHz (right panel), with contours traced at $(-1,1,2,4,\dots,32)\times$ the $\sim 3 \sigma$ noise level, which is 0.2 and 0.6 mJy\,beam$^{-1}$ at 5 and 8 GHz, respectively. The half peak beam width is shown in the upper left corner of each panel, and is 4.6 mas $\times$ 1.4 mas in position angle (PA) $-19^\circ$ and 8.0 mas $\times$ 3.4 mas in PA $19^\circ$ at 5 and 8 GHz, respectively. Note the unusually larger beam at 8 GHz, due to the lack of fringe detections on the longest baselines at this frequency. In the 5 GHz image, model-fit elliptical Gaussian components are shown as crosses and labelled E, B, W, from east to west. \label{f.vlba}}
\end{figure*}

\section{Introduction}

Ionized gas outflows from the nuclei of active galaxies (active galactic nuclei, hereafter AGN) are a relatively common phenomenon. They are of great interest since they transport mass and energy to the host galaxy in a process known as AGN feedback, which has fundamental impact on galaxy properties and evolution \citep{DiMatteo2005}. In this context, ultra-fast outflows \citep[UFOs, see e.g.\ ][]{Chartas2002,Pounds2003} are particularly relevant because of their mass outflow rate ($0.01-1 M_\odot$ yr$^{-1}$) and kinetic energy ($10^{42}-10^{45}$ erg\, s$^{-1}$). A much less common phenomenon (less than 10\% of AGNs) is the presence of a pair of relativistic jets of non-thermal plasma emerging from the central nucleus, which are generally detected in the form of collimated structures at radio wavelength. In recent years, several works have studied the coexistence of fast outflows and relativistic jets and their possible interplay \citep[see ][and references therein]{Tombesi2014}. A connection between the two phenomena is also predicted by magneto-hydrodynamic (MHD) models \citep[e.g.][]{Tchekhovskoy2011,Fukumura2014}.

\iras\ (located at R.A. 17h\,03m\,30.3830s, Dec.\ +45$^\circ$\, 40\arcmin\, 47\arcsec.167, $z=0.0604$) is an ideal target to further explore this subject. It is a bright Narrow Line Seyfert 1 (NLS1) spiral galaxy with black hole mass of $M_\mathrm{BH}\sim 5.9\times10^6 M_\odot$ \citep{Wang2001}. We have  observed it in January 2014 with \xmm, obtaining a high-resolution X-ray spectrum. Our observations reveal a series of absorption lines corresponding to at least 5 outflowing components with velocity in the range of  23,000-33,000 km/s \citep{Longinotti2015}. The energetics of the wind indicate that it might be able to produce feedback on the host galaxy.  Evidence for such phenomenon is also provided by the presence of  large scale molecular gas outflowing at low velocity that is associated to the fast X-ray wind (Longinotti et al. submitted), as predicted by several feedback models (e.g. Faucher-Giguere \& Quataert 2012, Zubovas \& King 2012). At the same time, the presence of bright and compact radio emission indicates that this source is a radio loud (RL) NLS1 and that it also has a radio jet \citep{Doi2007,Doi2011,Gu2010}.

In the present paper, we present new high angular resolution, multi-frequency Very Long Baseline Interferometry (VLBI) radio observations obtained near simultaneously with the XMM-{\it Newton} observations. We further report on the analysis of archival VLBI datasets, to provide constraints on the properties of the ejection process in this source. 

The paper is organized as follows: we describe our new observations and archival data analysis in Sect.~\ref{s.3}; we present the results in Sect.~\ref{s.4} and a review of the radio properties of \iras\ from the literature in Sect.~\ref{s.2}; finally, we discuss them in Sect.~\ref{s.5}. Throughout the paper, we use a $\Lambda$CDM cosmology with $h = 0.705$, $\Omega_m = 0.27$, and $\Omega_\Lambda=0.73$ \citep{Komatsu2009}. With these values, the redshift of the source corresponds to a luminosity distance of $D_L = 269$ Mpc and to an angular scale of 1\,mas\,=\,1.16\,pc; proper motion of 1 mas/yr corresponds to an apparent speed of $4.0\,c$. The spectral index $\alpha$ is defined such that $S(\nu)\sim\nu^{-\alpha}$ and position angles (P.A.) are defined positive north through east.

\section{Observations and data reduction}\label{s.3}

\subsection{New data}
We observed \iras\ with the NRAO Very Long Baseline Array (VLBA) on 2014 February 12, 20 days later than the \xmm\ pointing. We observed the source  at 5, 8, 15, and 24 GHz, for a net integration time of 10, 20, 60, and 60 minutes, respectively. We recorded data with the new 2 Gbps rate offered by the VLBA, corresponding to 8 baseband channels of 32 MHz bandwidth each, in full polarization. Based on the high sensitivity provided by this observing mode and on the flux density published by \citet{Gu2010}, we did not observe a phase calibrator for phase referencing.

We carried out the data reduction following the usual procedures in AIPS, applying calibrations for the total electronic content of the ionosphere at $\nu \le 8$ GHz, the earth orientation parameters, the correlator digital sampling, the combined system temperature and gain curves, the parallactic angle, and the instrumental single band delays. We then fringe fitted the data in order to remove residual delays, rates and phase variations; we used as input model an image produced from the same dataset analysed by \citet{Gu2010} but re-analysed by ourselves (see Sect.~\ref{s.gu}). At 5 GHz, we found solutions for most stations (all but Brewster); at 8 GHz we found solutions only for stations forming relatively short baselines; at 15 and 24 GHz, we could not find any solution. For the 5 and 8 GHz datasets, we then proceeded with the standard iterations of hybrid mapping, with phase only self-calibration at first and phase and amplitude self-calibration in the final cycles. 

\subsection{Archival data} \label{s.gu}

\subsubsection{2000 June \& August data}
The peak brightness and total flux density at 5 GHz obtained in our observations are $\sim 5$ times lower than those reported at the same frequency by \citet{Gu2010}. Considering the non-detections at higher frequency, we are confident that the error on our measurement is not larger than the typical, conservative 10\% calibration uncertainty. Therefore, in order to investigate possible amplitude and structural variability we downloaded and re-analysed the VLBA datasets presented by \citet{Gu2010}. These datasets are BM033D and BM033E, taken on 2000 June 16 and August 21, respectively. Both datasets were obtained with 2 IFs of 8 MHz bandwidth each, in dual polarization, for a total data rate of 128 Mbps. The delay, rate, and phase solutions were determined on the nearby calibrator J1713+4916 and transferred to the target source; other steps of the calibration were identical to those described for the new observations. The total integration time on \iras\ was about 100 minutes per observation.

The images at the two 2000 epochs are consistent in terms of intensity and structure, so we combined the two visibility datasets to further improve the image sensitivity and fidelity. We then produced a final image with an additional iteration of phase only and phase and amplitude self-calibration.

\subsubsection{2012 Jan \& Feb data}

We also found and downloaded two five minute scans on \iras\ obtained in the context of a large observational VLBA project at 8 GHz. These observations took place on 2012 Jan 12 (VLBA experiment code: BC196ZQ) and 2012 Feb 8 (code BC201AC). In both experiments, 8 channels with 16 MHz bandwidth each and single polarization were recorded and correlated. 
The source was bracketed by 1 minute scans on 1705+456, which we used to determine delay, rate, and phase solutions. Hancock and Saint-Croix did not produce data in either run because the calibrator was not detected (probably resolved out) on baselines to these stations. After the standard calibration, we split the data without averaging in time nor frequency (other than within each of the eight IFs), since the phase tracking position was several hundreds milliarcsecond from the position of the source determined in the other phase referencing experiment. In this way, we were able to produce images not affected by smearing and indeed we detected the source in both datasets. 

\begin{figure}
\includegraphics[width=0.96\columnwidth]{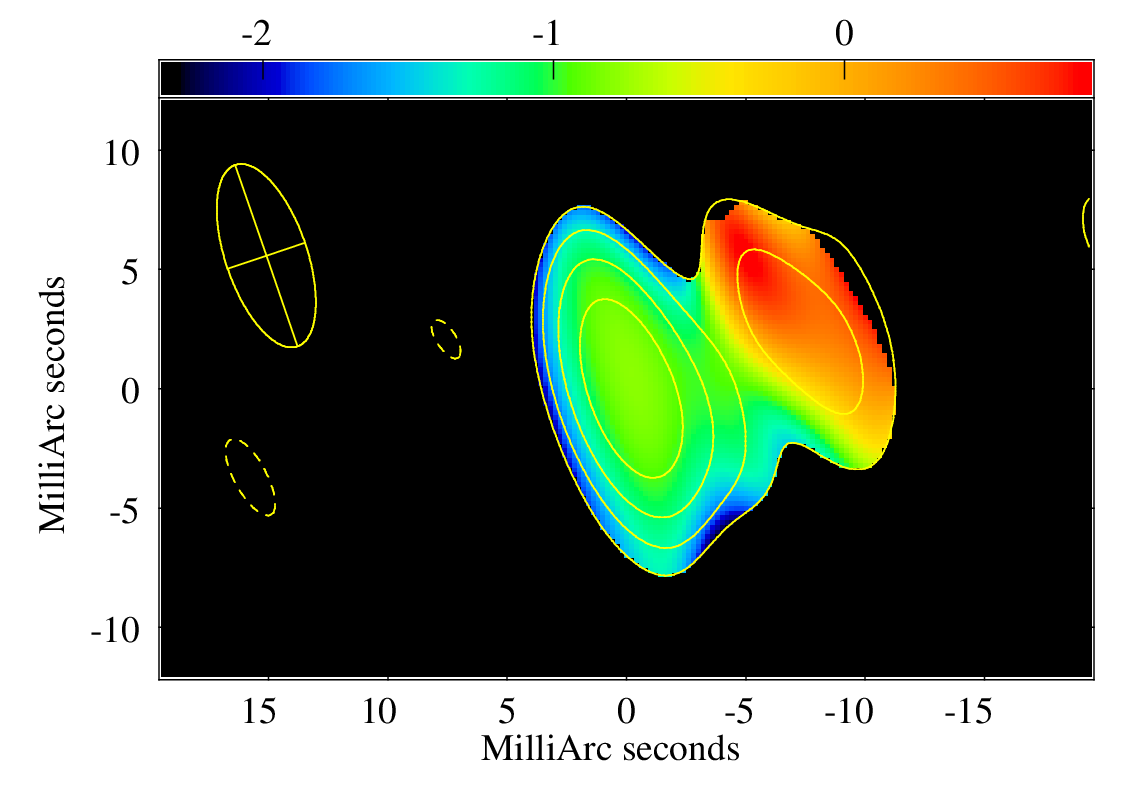}
\caption{Simultaneous 5-8 GHz spectral index VLBA image of \iras\ on 2014 Feb 12. Contours show total intensity at 5 GHz, convolved with the 8 GHz restoring beam; colours show the spectral index values. \label{f.spix}}
\end{figure}

\section{Results}\label{s.4}

\subsection{New data}

We show in Fig.~\ref{f.vlba} the final images at 5 and 8 GHz. At 5 GHz, the source is extended in east-west direction, with a total flux density of about 23 mJy. A compact component with peak brightness 8.9 mJy\, beam$^{-1}$ is surrounded by diffuse emission on both sides, brighter and more extended on the western end. At 8 GHz, the overall morphology is similar. However, due to the lack of visibilities on the longest baselines the final sensitivity and angular resolution are poorer than at 5 GHz. The source can then be most basically described as a compact asymmetric double, with a brighter eastern component (9.0 mJy\, beam$^{-1}$ peak brightness) and a weaker, somewhat more diffuse western feature. There is clearly also some extended emission bridging the two features. The total flux density at 8 GHz is 13 mJy.

We carried out a model-fit in the image plane, in order to measure the brightness temperature of the main component, i.e.\ the eastern component labelled {\it E}. The brightness temperature is obtained from the following formula:

$$T_b=\frac{1}{2k_B} \left( \frac{c}{\nu} \right)^2 \frac{S(1+z)}{\pi a b}$$

where $k_B$ is the Boltzmann constant, $c$ is the speed of light, $\nu$ is the observing frequency, $S$ is the total flux density of the component, $a$ and $b$ are the major and minor semi-axis of the component. In our case, $\nu=5.0$ GHz, $z=0.0604$ and, for component $E$,  $S=14$ mJy, $a=2.5$ mas, $b=1.0$ mas. The resulting brightness temperature is $T_b=1.0\times10^8$ K.

We show in Fig.~\ref{f.spix} the first quasi-simultaneous spatially resolved spectral index image of the source, based on our 5 and 8 GHz data. Before combining the images, we have convolved the 5 GHz image with the same restoring beam of the 8 GHz data, and clipped all the pixels below the $3\sigma$ noise level in either image. As a result of this convolution, the compact eastern feature becomes ``contaminated'' by some additional extended emission. The image then shows a steep spectrum integrated over the eastern component and a flatter one in the western side. However, the uncertainty on the western side spectral index is larger, and the eastern component is certainly affected by the presence of diffuse emission, which likely bias the results to steeper values. Indeed, if we only take the peak brightness of the component we find similar values, indicating a flat spectrum for the most compact emission region. Considering the emission detected in the entire VLBI images, the simultaneous spectral index is $\alpha=1.0\pm0.4$.

Finally, we estimate the upper limit to the brightness of any compact structure, based on the non detection of the source at 24 and 15 GHz, and on the longest baselines at 8 GHz. The $5\sigma$ VLBA sensitivity for a single baseline, channel, and polarization at 8, 15, and 24 GHz is 12, 28, and 33 mJy\,beam$^{-1}$, respectively. This assumes 10, 5, and 5 minutes integration times, respectively, for a structure with angular size of 4 mas at 8 GHz (based on the non detection on baselines longer than 45 M$\lambda$), 17 mas at 15 GHz and 11 mas at 24 GHz (based on the non detection on any baselines in the 15 and 24 GHz datasets). These correspond to upper limits on the brightness temperature of $2.2\times10^7$ K, $1.5\times10^7$ K, and $6.7\times10^6$ K, at the three frequencies, respectively.

\subsection{Archival data}

\begin{figure}
\includegraphics[width=0.96\columnwidth]{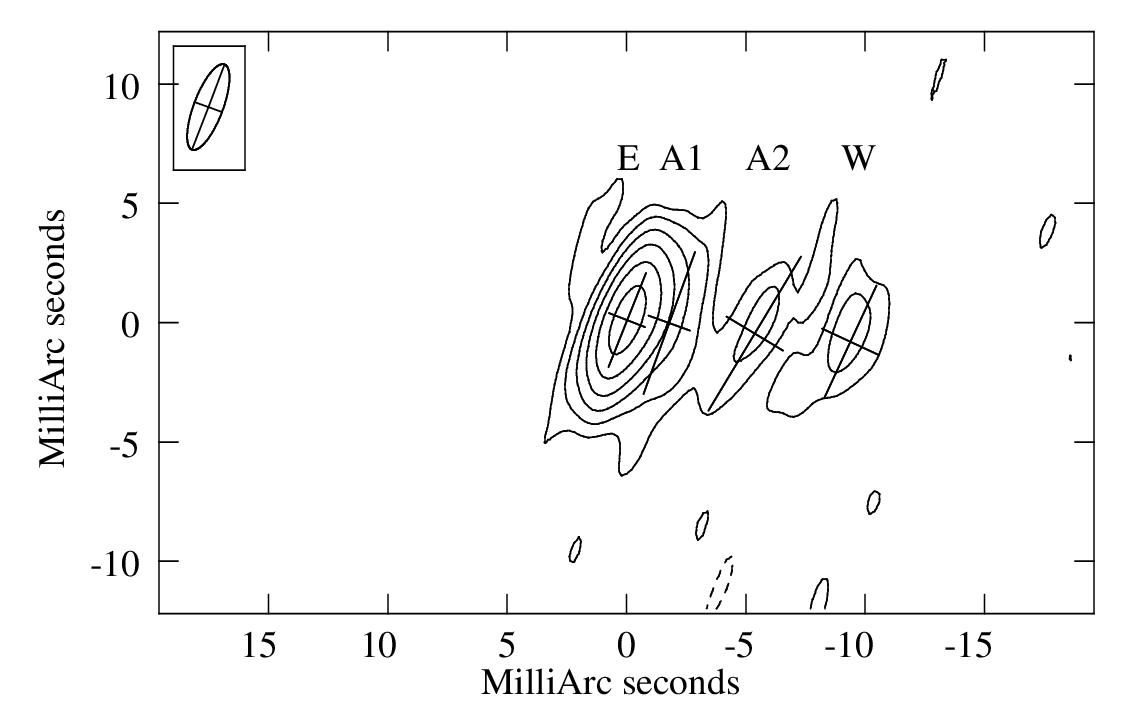}
\caption{VLBA image of \iras\ on the combined dataset from 2000 June and August observations at 5 GHz, with contours traced at $(-1,1,2,4,\dots,32)\times$ the $\sim 3 \sigma$ noise level, which is 0.27 mJy\,beam$^{-1}$. The half peak beam width is shown in the upper left corner of each panel and it is 3.8 mas $\times$ 1.2 mas in PA $-21^\circ$. Crosses and letters indicate and label model-fit elliptical Gaussian components. \label{f.2000}}
\end{figure}

We show in Fig.~\ref{f.2000} the image obtained from the reanalysis of the combined dataset obtained from the 2000 observations. The morphology is overall consistent with both our new observations and the images reported by \citet{Gu2010}: the brightest component is located at the eastern end, surrounded by  diffuse emission more extended in the western direction. Local peaks are embedded in this emission, with a final component located about 9 mas west of the image peak. The total flux density and the peak brightness recovered ($S_5 \sim 25$ mJy and $P_5 \sim 13$ mJy\,beam$^{-1}$, respectively) are in much closer agreement with the 2014 data than with the values reported by \citet{Gu2010}. 

We tried to investigate on the origin of this discrepancy but could not come up with a final explanation. We determined a flux density of $\sim80$ mJy for the phase calibrator, which is lower than the 2.3 and 8 GHz values reported in the VLBA calibrator list (148 mJy on short spacings and 117 mJy on long spacings at 2.3 GHz, 217 and 182 mJy at 8 GHz); this might have prompted \citet{Gu2010} to rescale the amplitudes up by a factor of a few, resulting in the high flux density for the target. However, the flux density of 3C345 observed in the same experiment did not reveal any sign of being underestimated, and later 5 GHz observations of J1713+4916 found flux density values closer to our measurement; considering also a fair amount of variability typical for compact sources, we do not see any reason to modify the amplitude scale for the 2000 observations of \iras.

\begin{figure}
\includegraphics[width=0.96\columnwidth]{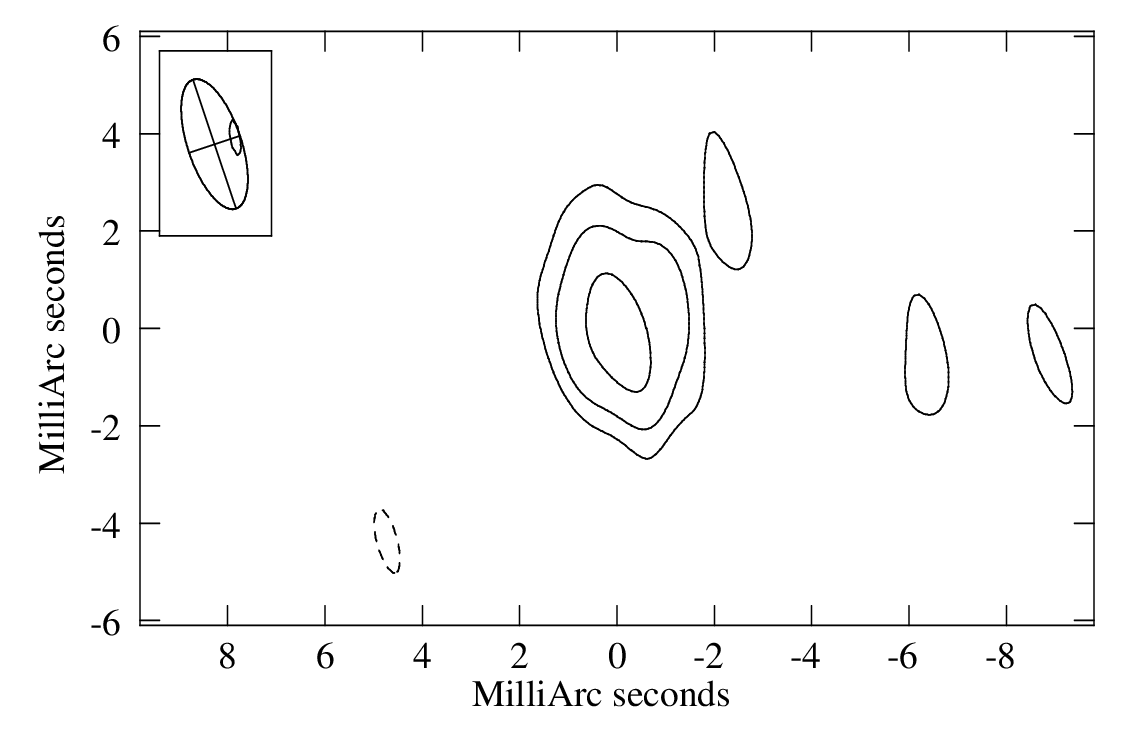}
\caption{VLBA 8 GHz image of \iras\ on 2012 Feb 8, with contours traced at $(-1,1,2,4)\times$ the $\sim 3 \sigma$ noise level, which is 1.0 mJy\,beam$^{-1}$. The half peak beam width is shown in the upper left corner of each panel and it is 2.8 mas $\times$ 1.1 mas in PA $18^\circ$. \label{f.2012}}
\end{figure}

In the 2012 8 GHz experiments, we detect a slightly resolved component, with an extension to the north-western side and faint diffuse emission to the west (see Fig.~\ref{f.2012}). The peak brightness is 7.4 and 5.9 mJy\,beam$^{-1}$ in 2012 Jan and Feb, respectively. Given the very short observation time, we can not make strong claims about the image fidelity and any significant variability. In any case, the total cleaned flux density is consistent among the epochs with a value of $S_8=11.5$ mJy, also similar to the result of our 2014 observations.

The astrometry of the 2000 and 2012 observations is not entirely consistent, with a $\sim 10$ mas offset between the core peak at the two frequencies. The 8 GHz observation positions are in good agreement (within $<1$ mas) with those reported at the same frequency by \citet{Doi2007}.

\begin{table}
\begin{center}
\begin{tabular}{llccc}
\hline
Epoch & Component & $x$ & $y$ & $S_5$ \\
(year) & & (mas) & (mas) & (mJy)  \\
\hline
2000 & E & 0.0 & 0.0 &  17.9 \\
 & A1 & $-1.3$ & $-0.13$ & 3.5 \\
 & A2 & $-3.9$ & $-0.6$ & 2.4 \\
 & W & $-9.4$ & $-0.9$ & 2.0 \\
2014 & E & 0.08	 & $-0.05$ & 14.1 \\
 & B & $-3.1$ & $-1.9$ & 	5.2 \\
 & W & $-9.1$ & $-0.8$ & 	1.5 \\
\hline
\end{tabular}
\end{center}
\caption{Results of model-fit to the 5 GHz images of \iras. The coordinates are referenced to the peak position of the brightest component in the 2000 combined images.}\label{t.modelfit}
\end{table}

\subsection{Model fits and proper motion}

We report in Table~\ref{t.modelfit} the results of a model-fit in the image plane with elliptical Gaussian components to the 2000 and 2014 5~GHz images. We use four components to describe the 2000 data (which have better coverage of the $(u,v)$-plane) and three components for the 2014 data. In any case, the main features at the two edges of the structure can be easily identified. Their positions are consistent to better than one beam and the flux densities are also consistent within the calibration uncertainty.

The separation between $E$ and $W$ has changed from 9.4 mas in 2002 to 9.2 mas in 2014; since the difference is less than $1/5$ of the beam, we claim that both components are stationary, and that their separation velocity must be smaller than 0.015 mas\,yr$^{-1}$, or $0.06c$.

\section{Radio properties of \iras}\label{s.2}

\iras\ is also known as B3\,1702+457, since it was detected in the B3 survey at 408 MHz with the Bologna Northern Cross \citep{Ficarra1985}. 
As it is a B3 source, it is quite bright and indeed it is detected in many other surveys: the 6th Cambridge survey of radio sources \citep[6C,][]{Hales1988}, the Texas survey of radio sources \citep{Douglas1996}, and the Westerbork Northern Sky Survey \citep[WENSS,][]{Rengelink1997} at low frequency; the NRAO VLA Sky Survey \citep[NVSS,][]{Condon1998} and the Faint Images of the Radio Sky at Twenty Centimeters \citep[FIRST,][]{Becker1995} at 1.4 GHz; the 87GB and GB6 at 5 GHz \citep{Gregory1991,Gregory1996}. From its flux density and luminosity distance, the radio power of \iras\ is $P_\mathrm{150 \ MHz}=3.2 \times 10^{24}$~\whz ($\nu L_\nu=4.8 \times 10^{39}$~\ergs) and $P_\mathrm{1.4 \ GHz}=1.0 \times 10^{24}$~\whz ($\nu L_\nu=1.5 \times 10^{40}$~\ergs), calculated at 150 MHz from the 6C and at 1.4 GHz from the NVSS values, respectively.

\begin{figure}
\includegraphics[width=0.96\columnwidth]{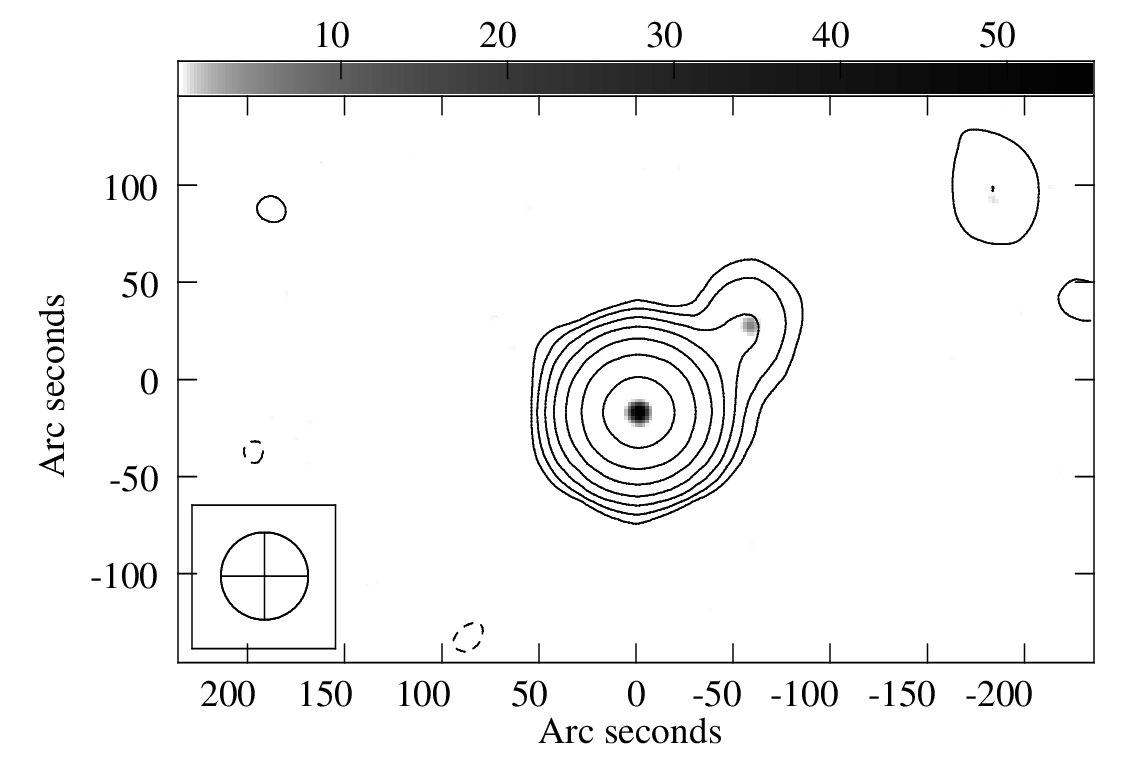}
\caption{Overlay of the NVSS (contours) and FIRST (grey scale) 1.4 GHz VLA images. The NVSS HPBW is 45\arcsec\ and is shown in the bottom left corner; the FIRST HPBW is 4.5\arcsec. Contours are traced at $(-1,1,2,4,\dots)\times 1.3$~mJy\,beam$^{-1}$; the grey scale is shown in the top wedge and spans the range between 0.4 and 55~mJy\,beam$^{-1}$ . The half peak beam width is shown in the upper left corner of each panel and it is 2.8 mas $\times$ 1.1 mas in PA $18^\circ$. \label{f.firstnvss}}
\end{figure}

The FIRST image (shown with grey tones in Fig.~\ref{f.firstnvss}) reveals one main, bright, compact component, unresolved at the $\sim5^{\prime\prime}$ angular scale, and a secondary, fainter component located at 1.2\arcmin\ (81 kpc projected at the \iras\ redshift), in PA $-52^\circ$, which is also visible in the NVSS image (contours in Fig.~\ref{f.firstnvss}) as an extension of the main component. The flux density in the NVSS is only slightly larger than in the FIRST (121 vs 118 mJy), indicating that the source is compact at this angular resolution; the total flux density in the NVSS might also be slightly affected by the secondary component, which is only partly resolved at the 45\arcsec\ angular resolution of the survey. The two components are in any case well separated in the FIRST, which can be used to determine their brightness ratio ($118:6$, or $\sim20$). We are thus confident that any flux density measured at lower resolution is only slightly affected by the presence of this component. A visual inspection of the SDSS images does not reveal any obvious optical counterpart, so it remains uncertain whether this is a background object or a large-scale feature associated to \iras\ itself. We note that kiloparsec scale emission is not uncommon in NLS1s, having been discovered in 10 objects so far \citep{Anton2008,Doi2012,Doi2015,Gliozzi2010,Richards2015,Whalen2006}. However, with $\sim 80$ kpc (projected) linear size, \iras\ would be the largest NLS1 known so far. If the third weaker component located further north-west were also physically associated to \iras, the size would become even larger. Deep, low frequency observations with a few arcsecond angular resolution will be necessary to clarify this issue.

In addition to these 1.4 GHz data, \citet{Snellen2004} considered a number of additional multi-frequency interferometric and single dish datasets between 150 MHz and 22 GHz. They report a  spectral index of $\alpha=-0.88$, slightly flattening at low frequency, and provide constraints on the turnover frequency and peak flux density of $\nu_\mathrm{peak}<150$ MHz and $S_\mathrm{peak}>380$ mJy, respectively. A low frequency turnover is confirmed by the flatter ($\alpha=-0.23$) spectral index between 325 and 1400 MHz reported by \citet{Massaro2014} and the non detection at 74 MHz in the VLA Low-Frequency Survey \citep[VLSS, whose typical point-source detection limit is 0.7 Jy beam$^{-1}$,][]{Cohen2007}. On the basis of its spectrum and linear size,  \citet{Snellen2004} classify the source as a candidate low power compact steep spectrum (CSS) source and include it in the ``compact radio sources at low redshift'' (CORALZ) sample\footnote{However, later high angular resolution observations of the full sample \citep{de Vries2009} do not include the source, without any reason given in the paper}.

To construct their spectra, \citet{Snellen2004} considered data taken at the same frequency in different epochs, thus providing us also variability information: only small deviations are present, suggesting that variability in the radio is not significant; if we take into account possible calibration uncertainties and different angular resolution in the various observations, the amount of real source variability has to be very limited, if any. In this sense, the most remarkable finding is the comparison of the single dish (87GB and GB6) and interferometric (VLA) 5 GHz flux density values, with $S_\mathrm{87GB} = 26\pm6$ mJy, $S_\mathrm{GB6} = 29\pm4$ mJy, and $S_\mathrm{VLA} = 40\pm4 $ mJy. It is not clear how reliable are the single dish measurements, since they are near the flux density limit of the catalogs (25 and 18 mJy for the 87GB and the GB6, respectively). As the 87GB data were reported by many authors to estimate the radio loudness ($R$), if the real flux density is larger than 26 mJy, then also the real $R$ could be accordingly higher by $\Delta \log R \sim +0.2$. On the issue of radio loudness, it is interesting to note that the values of $R$ reported in the literature vary significantly in different works, typically based on the assumptions on the optical spectrum used to estimate the $B$ magnitude. The most recent value is $\log R = 2.17$ \citep{Doi2016}, based on the ratio of FIRST 1.4 GHz radio to SDSS $g$-band. Such values puts \iras\ firmly in the radio loud regime.

At high angular resolution, we can compare our new and reprocessed VLBA data with other VLBI observations, in particular the 8.4 GHz Japanese VLBI Network (JVN) observations by \citet{Doi2007} and the 1.6 VLBA data by \citet{Doi2011}. These two datasets show a fairly resolved morphology, which accounts for over one half of the total flux density measured at the same frequency by, e.g., the VLA. In particular, $S_{1.6} = 79.9\pm4.0$ mJy and $S_{8.4} = 18.5\pm2.6$ mJy. The resulting non simultaneous spectral index is $\alpha = 0.84\pm0.21$, consistent with the one found with the simultaneous data. A further non detection at 22 GHz at the $<9$ mJy level, entirely consistent with our result, is also reported by \citet{Doi2016}.

\section{Discussion}\label{s.5}

Based on the information gathered by the new, the archival, and the literature data on \iras, we determine that it is a jetted non-thermal source. The steep overall spectrum on $\sim$ kpc \citep[$\alpha=0.9$,][]{Snellen2004} and $\sim$ pc ($\alpha=1.0$, this work) scales indicates that the radio emission has a synchrotron origin. The total radio luminosity ($P_\mathrm{1.4 \ GHz}=1.0 \times 10^{24}$~\whz) and, above all, the brightness temperature of the VLBI main component ($T_b=1.0\times10^8$ K) are far too large for starburst processes, and constrain the radio emission to have  predominantly a nuclear origin. The difference between the VLBI and total flux density reveals the presence of radio emission on pc-to-kpc scales, most likely fuelled by the central region. Taken together, these elements indicate that the black hole in \iras\ is powering a non-thermal plasma jet, in addition to the  X-ray outflow of gas studied by \citet{Longinotti2015}. 

From this starting point, we can then can try to classify the source in a few different ways, complementary rather than conflicting. As already proposed by \citet{Snellen2004}, the compact structure and steep integrated radio spectrum suggest that \iras\ is a candidate CSS source. Differently from compact flat-spectrum sources, such as blazars, CSS sources are oriented on the plane of the sky, so their compactness is intrinsic rather than due to projection effects, as highlighted by the lack of a prominent compact flat-spectrum core. CSS sources are thought to be the early stages of evolution of the more extended, classical FR2 radio galaxies, as indicated by their young kinematic and spectral age estimates \citep{Fanti1995,Murgia1999,Giroletti2009b}. The parsec scale structure presented in this work is in agreement with the lack of a prominent core and of strong beaming effects; however, the absence of component motion over a 14 years time scale does not indicate that this source is in a state of rapid growth, which would be typical of CSS sources. Moreover, the radio luminosity of \iras\ is also significantly ($\sim10^{-4}\times$) lower than that of the archetypical CSS sources, suggesting that it could lie in a different evolutionary path.

\citet{Giroletti2005} and \citet{Kunert-Bajraszewska2010} have indeed introduced the class of low-power compact (LPC) or low-luminosity compact (LLC) objects, which similarly to the CSS sources are thought to be intrinsically compact, yet they present lower luminosities ($P_{1.4\ \mathrm{GHz}}<10^{26}$ W Hz$^{-1}$) and slower or null advance velocities. These sources might be short-lived objects or the progenitors of low luminosity FR1 radio sources. In terms of radio properties (morphology, spectrum, power), \iras\ fits well within this class; however, its peculiar optical properties do not match the typical features of a FR1 progenitor: the host is a barred spiral galaxy \citep{Ohta2007} and the optical spectrum characterizes the source as a NLS1.

The presence of radio emission in NLS1 has attracted a lot of interest in recent times, especially after the discovery of gamma-ray emission from a growing number of RL NLS1s \citep{Abdo2009a,Abdo2009b,D'Ammando2012,D'Ammando2015}. Gamma-ray emission in NLS1s is ascribed to the presence of a beamed relativistic jet, similar to what happens in blazars, although some similarities with CSS sources in spectrum and size have been claimed, e.g.\ in PKS 2004-447 \citep{Gallo2006,Orienti2015,Schulz2015} and SDSS\,J143244.91+301435.3 \citep{Caccianiga2014}. Evidence for beaming in RL NLS1s in general (not only in gamma-ray NLS1s) has been found in several objects, in the form of dominant cores, high brightness temperatures \citep[above $10^{10}$ K, e.g.][]{Giroletti2011,Wajima2014,Gu2015}, and occasionally the presence of superluminal motion \citep{D'Ammando2012}. Our results on the morphology and brightness temperature indicate that \iras\ is different from this sub-set of the NLS1 population; yet, it remains also markedly separated from the bulk of the radio quiet NLS1 population, which have much lower radio powers (as low as $P_{1.4\ \mathrm{GHz}}\sim10^{19}$~\whz) and are seldom detected with VLBI \citep{Giroletti2009a,Doi2013}.  One intriguing possibility is that \iras\ belongs to the elusive population of misaligned RL NLS1s \citep{Berton2015}.

From a multi-wavelength perspective,  \iras\ is also highly remarkable for its X-ray properties. While its X-ray and bolometric luminosities are not exceptional ($L_\mathrm{0.3-10\ keV} \sim 1.5 \times 10^{44}$ \ergs, $L_\mathrm{BOL} \sim 5.2 \times 10^{44}$ \ergs), the recent high spectral resolution ($E/\Delta E \sim 1000$) XMM-Newton observations of \citet{Longinotti2015}, obtained within one month of our VLBA data, have revealed a unique sub-relativistic outflow. The spectrum shows a series of absorption lines corresponding to at least five absorption components with velocities in the range of 23,000--33,000 \kms; the associated column densities observed in the X-ray wind span  from $5\times 10^{20}$ to $\sim10^{24}$ cm$^{-2}$. The mass and energy output estimated by these authors indicate that the higher column density wind is capable of expelling sufficient gas to produce feedback onto the host galaxy, even if the source is $\sim100$ times less luminous than quasars where this phenomenon is commonly observed \citep{Nardini2015,Tombesi2015}. Whether the presence of radio emission is somehow related to this peculiar wind is a question worth exploring. In a similar case of coexistence of fast X-ray outflow and radio jet, albeit in a high radio power source, \citet{Tombesi2012} reported that the  pressure estimated by the X-ray column density is comparable to the pressure of the radio inner jet, therefore lending support to the idea that the X-ray outflow may help to collimate the jet. In the case of \iras\, higher resolution VLBA observations will provide a better estimate of the spatial scale of the jetted structure and its properties, shedding light on the possible wind-jet connection. 

The mutual interaction between AGN winds and jets is the subject of much debate. \citet{Zakamska2014} found a correlation between radio power and gas kinematics (in particular, [OIII] line velocity width) in a sample of luminous type-2 radio quiet quasars. They considered several scenarios to interpret this correlation, and concluded that either (i) radiatively driven winds propagate into the interstellar medium of the host galaxy, generating shock fronts which in turn accelerate particles producing radio emission or (ii) mechanical energy of relativistic jets heats overpressured cocoons, which then launch winds of ionised gas \citep{Mullaney2013}. The detection of the X-ray outflow in \iras\ and the source proximity would in principle permit us to study this issue with adequate linear resolution. However, the radio luminosity $\log \nu L_\nu = 40.2$ ([erg\,s$^{-1}$], calculated at $\nu=1.4$\,GHz) is comparatively large in comparison to the bolometric luminosity, confirming that \iras\ is a radio-loud source and therefore quite exceptional with respect to the radio-quiet population studied by \citet{Zakamska2014}.

\section{Summary}
\begin{itemize}

\item The radio luminosity ($\log L_{150}=24.5$ W Hz$^{-1}$) is typical of FR1 radio galaxy, but the structure is compact, i.e.\ $<$ kpc scale. As already hinted at by other works \citep{Snellen2004}, it meets the criteria for a low power compact source \citep[LPC,][]{Giroletti2005}.

\item The integrated steep spectrum ($\alpha=-0.9$) clearly indicates that the radio emission has a synchrotron origin, i.e.\ the presence of relativistic particles and magnetic fields.

\item On parsec scales, the images at 5 and 8 GHz indicate that the source is clearly extended in an amorphous yet predominantly linear structure, indicating the presence of a jet or outflow on scales of about 10 parsecs; also the non detections at 15 and 24 GHz, when compared with the VLA flux densities at the same frequency, indicate that the source structure is heavily resolved.

\item Within these diffuse emission, there is a main, brighter component, which one would naturally identify with the ``core'', which however is not characterized by a classical flat/inverted spectrum.

\item The lack of a prominent flat spectrum component suggests that the amount of beaming in any possible core is negligible, i.e.\ the plasma does not have a bulk relativistic motion or it is moving at a very large angle to the line of sight - also in agreement with the lack of variability.

\item The lack of motion over 15 years suggests that the separation speed is lower than 0.2 mas/14 years, i.e.\ $<0.06c$. 

\end{itemize}

All these properties locate \iras\ as a unique object, not easily classifiable as any standard radio emitting AGN.  In addition \iras\ is an optimal laboratory to study in detail AGN feedback. The coexistence of an ultrafast X-ray wind and a molecular large scale outflow indicate that this process is in action via an induced shock. The peculiar radio properties  are likely related to this phenomenon. Future radio observations up to 24 GHz in phase reference mode will allow us to obtain higher fidelity total intensity and spectral images, useful to investigate the role of the radio emission  in this context.

\begin{acknowledgements}  
We acknowledge financial contribution from grant PRIN-INAF-2014. This research has made use of NASA's Astrophysics Data System Bibliographic Services. This research has made use of the NASA/IPAC Extragalactic Database (NED) which is operated by the Jet Propulsion Laboratory, California Institute of Technology, under contract with the National Aeronautics and Space Administration. The National Radio Astronomy Observatory is a facility of the National Science Foundation operated under cooperative agreement by Associated Universities, Inc.

\end{acknowledgements}


\bigskip \noindent
\begin{thebibliography}{}

\bibitem[Abdo et al.(2009a)]{Abdo2009a} Abdo, A.~A., Ackermann, M., Ajello, 
M., et al.\ 2009a, ApJ, 699, 976 

\bibitem[Abdo et al.(2009b)]{Abdo2009b} Abdo, A.~A., Ackermann, M., Ajello, 
M., et al.\ 2009b, ApJ, 707, L142 

\bibitem[Ant{\'o}n et al.(2008)]{Anton2008} Ant{\'o}n, S., Browne, 
I.~W.~A., \& March{\~a}, M.~J.\ 2008, A\&A, 490, 583 

\bibitem[Becker et al.(1995)]{Becker1995} Becker, R.~H., White, R.~L., 
\& Helfand, D.~J.\ 1995, ApJ, 450, 559 

\bibitem[Berton et al.(2015)]{Berton2015} Berton, M., Foschini, L., Ciroi, 
S., et al.\ 2015, A\&A, 578, A28 

\bibitem[Caccianiga et al.(2014)]{Caccianiga2014} Caccianiga, A., 
Ant{\'o}n, S., Ballo, L., et al.\ 2014, MNRAS, 441, 172 

\bibitem[Chartas et al.(2002)]{Chartas2002} Chartas, G., Brandt, W.~N., 
Gallagher, S.~C., \& Garmire, G.~P.\ 2002, ApJ, 579, 169 

\bibitem[Cohen et al.(2007)]{Cohen2007} Cohen, A.~S., Lane, W.~M., Cotton, 
W.~D., et al.\ 2007, AJ, 134, 1245 

\bibitem[Condon et al.(1998)]{Condon1998} Condon, J.~J., Cotton, W.~D., 
Greisen, E.~W., et al.\ 1998, AJ, 115, 1693 

\bibitem[D'Ammando et al.(2012)]{D'Ammando2012} D'Ammando, F., Orienti, M., 
Finke, J., et al.\ 2012, MNRAS, 426, 317 

\bibitem[D'Ammando et al.(2015)]{D'Ammando2015} D'Ammando, F., Orienti, M., 
Larsson, J., \& Giroletti, M.\ 2015, MNRAS, 452, 520 

\bibitem[de Vries et al.(2009)]{de Vries2009} de Vries, N., Snellen, 
I.~A.~G., Schilizzi, R.~T., Mack, K.-H., 
\& Kaiser, C.~R.\ 2009, A\&A, 498, 641 

\bibitem[Di Matteo et al.(2005)]{DiMatteo2005} Di Matteo, T., Springel, 
V., \& Hernquist, L.\ 2005, Natur, 433, 604 

\bibitem[Doi et al.(2011)]{Doi2011} Doi, A., Asada, K., 
\& Nagai, H.\ 2011, ApJ, 738, 126 

\bibitem[Doi et al.(2007)]{Doi2007} Doi, A., Fujisawa, K., Inoue, M., et 
al.\ 2007, PASJ, 59, 703 

\bibitem[Doi et al.(2012)]{Doi2012} Doi, A., Nagira, H., Kawakatu, N., 
Kino, M., Nagai, H., \& Asada, K.\ 2012, ApJ, 760, 41 

\bibitem[Doi et al.(2013)]{Doi2013} Doi, A., Asada, K., Fujisawa, K., 
Nagai, H., Hagiwara, Y., Wajima, K., \& Inoue, M.\ 2013, ApJ, 765, 69 

\bibitem[Doi et al.(2015)]{Doi2015} Doi, A., Wajima, K., Hagiwara, Y., 
\& Inoue, M.\ 2015, ApJ, 798, L30 

\bibitem[Doi et al.(2016)]{Doi2016} Doi, A., Oyama, T., Kono, Y., et al.\ 
2016, PASJ, 68, 73 

\bibitem[Douglas et al.(1996)]{Douglas1996} Douglas, J.~N., Bash, F.~N., 
Bozyan, F.~A., Torrence, G.~W., \& Wolfe, C.\ 1996, AJ, 111, 1945 

\bibitem[Fanti et al.(1995)]{Fanti1995} Fanti, C., Fanti, R., Dallacasa, 
D., Schilizzi, R.~T., Spencer, R.~E., 
\& Stanghellini, C.\ 1995, A\&A, 302, 317 

\bibitem[Faucher-Gigu{\`e}re \& Quataert(2012)]{2012MNRAS.425..605F} Faucher-Gigu{\`e}re, C.-A., \& Quataert, E.\ 2012, \mnras, 425, 605 

\bibitem[Ficarra et al.(1985)]{Ficarra1985} Ficarra, A., Grueff, G., 
\& Tomassetti, G.\ 1985, A\&AS, 59, 255 

\bibitem[Fukumura et al.(2014)]{Fukumura2014} Fukumura, K., Tombesi, F., 
Kazanas, D., et al.\ 2014, ApJ, 780, 120 

\bibitem[Gallo et al.(2006)]{Gallo2006} Gallo, L.~C., Edwards, P.~G., 
Ferrero, E., et al.\ 2006, MNRAS, 370, 245 

\bibitem[Giroletti et al.(2005)]{Giroletti2005} Giroletti, M., Giovannini, 
G., \& Taylor, G.~B.\ 2005, A\&A, 441, 89 

\bibitem[Giroletti 
\& Panessa(2009)]{Giroletti2009a} Giroletti, M., \& Panessa, F.\ 2009, ApJ, 706, L260 

\bibitem[Giroletti 
\& Polatidis(2009)]{Giroletti2009b} Giroletti, M., \& Polatidis, A.\ 2009, AN, 330, 193 

\bibitem[Giroletti et al.(2011)]{Giroletti2011} Giroletti, M., Paragi, Z., 
Bignall, H., et al.\ 2011, A\&A, 528, L11 

\bibitem[Gliozzi et al.(2010)]{Gliozzi2010} Gliozzi, M., Papadakis, I.~E., 
Grupe, D., Brinkmann, W.~P., Raeth, C., 
\& Kedziora-Chudczer, L.\ 2010, ApJ, 717, 1243 

\bibitem[Gregory 
\& Condon(1991)]{Gregory1991} Gregory, P.~C., \& Condon, J.~J.\ 1991, ApJS, 75, 1011 

\bibitem[Gregory et al.(1996)]{Gregory1996} Gregory, P.~C., Scott, W.~K., 
Douglas, K., \& Condon, J.~J.\ 1996, ApJS, 103, 427 

\bibitem[Gu \& Chen(2010)]{Gu2010} Gu, M., \& Chen, Y.\ 2010, AJ, 139, 2612 

\bibitem[Gu et al.(2015)]{Gu2015} Gu, M., Chen, Y., Komossa, S., Yuan, W., 
Shen, Z., Wajima, K., Zhou, H., \& Zensus, J.~A.\ 2015, ApJS, 221, 3 

\bibitem[Hales et al.(1988)]{Hales1988} Hales, S.~E.~G., Baldwin, J.~E., 
\& Warner, P.~J.\ 1988, MNRAS, 234, 919 

\bibitem[Longinotti et al.(2015)]{Longinotti2015} Longinotti, A.~L., 
Krongold, Y., Guainazzi, M., et al.\ 2015, ApJ, 813, L39 

\bibitem[Komatsu et al.(2009)]{Komatsu2009} Komatsu, E., Dunkley, J., 
Nolta, M.~R., et al.\ 2009, ApJS, 180, 330 

\bibitem[Kunert-Bajraszewska et al.(2010)]{Kunert-Bajraszewska2010} 
Kunert-Bajraszewska, M., Gawro{\'n}ski, M.~P., Labiano, A., 
\& Siemiginowska, A.\ 2010, MNRAS, 408, 2261 

\bibitem[Massaro et al.(2014)]{Massaro2014} Massaro, F., Giroletti, M., 
D'Abrusco, R., et al.\ 2014, ApJS, 213, 3 

\bibitem[Mullaney et al.(2013)]{Mullaney2013} Mullaney, J.~R., Alexander, 
D.~M., Fine, S., et al.\ 2013, MNRAS, 433, 622 

\bibitem[Murgia et al.(1999)]{Murgia1999} Murgia, M., Fanti, C., Fanti, R., 
Gregorini, L., Klein, U., Mack, K.-H., \& Vigotti, M.\ 1999, A\&A, 345, 769 

\bibitem[Nardini et al.(2015)]{Nardini2015} Nardini, E., Reeves, J.~N., 
Gofford, J., et al.\ 2015, Sci, 347, 860 

\bibitem[Ohta et al.(2007)]{Ohta2007} Ohta, K., Aoki, K., Kawaguchi, T., 
\& Kiuchi, G.\ 2007, ApJS, 169, 1 

\bibitem[Orienti et al.(2015)]{Orienti2015} Orienti, M., D'Ammando, F., 
Larsson, J., Finke, J., Giroletti, M., Dallacasa, D., Isacsson, T., 
\& Stoby Hoglund, J.\ 2015, MNRAS, 453, 4037 

\bibitem[Pounds et al.(2003)]{Pounds2003} Pounds, K.~A., Reeves, J.~N., 
King, A.~R., et al.\ 2003, MNRAS, 345, 705 

\bibitem[Rengelink et al.(1997)]{Rengelink1997} Rengelink, R.~B., Tang, Y., 
de Bruyn, A.~G., et al.\ 1997, A\&AS, 124,  

\bibitem[Richards 
\& Lister(2015)]{Richards2015} Richards, J.~L., \& Lister, M.~L.\ 2015, ApJ, 800, L8 

\bibitem[Schulz et al.(2015)]{Schulz2015} Schulz, R., Kreikenbohm, A., 
Kadler, M., et al.\ 2015, arXiv, arXiv:1511.02631 

\bibitem[Snellen et al.(2004)]{Snellen2004} Snellen, I.~A.~G., Mack, K.-H., 
Schilizzi, R.~T., \& Tschager, W.\ 2004, MNRAS, 348, 227 

\bibitem[Tchekhovskoy et al.(2011)]{Tchekhovskoy2011} Tchekhovskoy, A., 
Narayan, R., \& McKinney, J.~C.\ 2011, MNRAS, 418, L79 

\bibitem[Tombesi et al.(2012)]{Tombesi2012} Tombesi, F., Sambruna, R.~M., 
Marscher, A.~P., et al.\ 2012, MNRAS, 424, 754 

\bibitem[Tombesi et al.(2014)]{Tombesi2014} Tombesi, F., Tazaki, F., 
Mushotzky, R.~F., et al.\ 2014, MNRAS, 443, 2154 

\bibitem[Tombesi et al.(2015)]{Tombesi2015} Tombesi, F., Mel{\'e}ndez, M., 
Veilleux, S., et al.\ 2015, Nature, 519, 436 

\bibitem[Zubovas \& King(2012)]{2012ApJ...745L..34Z} Zubovas, K., \& King, A.\ 2012, \apjl, 745, L34

\bibitem[Wajima et al.(2014)]{Wajima2014} Wajima, K., Fujisawa, K., 
Hayashida, M., Isobe, N., Ishida, T., \& Yonekura, Y.\ 2014, ApJ, 781, 75 

\bibitem[Wang 
\& Lu(2001)]{Wang2001} Wang, T., \& Lu, Y.\ 2001, A\&A, 377, 52 

\bibitem[Whalen et al.(2006)]{Whalen2006} Whalen, D.~J., 
Laurent-Muehleisen, S.~A., Moran, E.~C., 
\& Becker, R.~H.\ 2006, AJ, 131, 1948 

\bibitem[Zakamska 
\& Greene(2014)]{Zakamska2014} Zakamska, N.~L., \& Greene, J.~E.\ 2014, MNRAS, 442, 784 

\end{thebibliography}
\end{document}